\documentstyle[aps,eqsecnum,epsf,preprint]{revtex}
\begin{document}
\def\Or {\overline{r}}
\def\Obr{\overline{\B.r}}
\def\Btensor{ B_{2n,\ell,\sigma}^{\alpha_1 \ldots \alpha_{2n}} }
\def\Ctensor #1#2{ \delta^{\alpha_{#1}\alpha_{#2}} B_{2n-2,\ell,\sigma}^
                    {\stackrel{ \text{no } #1 ,#2}
                    { \overbrace{_{\alpha_1 \ldots \alpha_{2n}} } }}}
\def\I.#1{\it #1}
\def\B.#1{{\bbox#1}}
\def\C.#1{{\cal #1}}
\def\BE{\begin{equation}}
\def\EE{\end{equation}}
\def\BEA{\begin{eqnarray}}
\def\EEA{\end{eqnarray}}

\title{{\rm DRAFT for PHYSICAL REVIEW E \hfill Version of \today }
\\~~\\
Nonperturbative Spectrum of Anomalous Scaling Exponents in the Anisotropic
Sectors of Passively Advected Magnetic Fields}
\author {Itai Arad$^*$, Luca Biferale$^{\dag}$ and Itamar Procaccia$^*$}
\address{$^*$Department of Chemical Physics,The Weizmann Institute of
Science, Rehovot
76100, Israel\\
$^{\dag}$ Dept. of Physics and INFM, University of Rome Tor Vergata \\
Via della Ricerca Scientifica 1, 00133 Italy, Rome}
\maketitle
\begin{abstract}
We address the scaling behavior of the covariance of the magnetic field in the
three-dimensional kinematic dynamo problem when the boundary conditions and/or
the external forcing are not isotropic. The velocity field is gaussian and
$\delta$-correlated in time, and its structure function scales with a positive
exponent $\xi$. The covariance of the magnetic field is naturally computed as
a sum of contributions proportional to the irreducible representations of the
SO(3) symmetry group. The amplitudes are non-universal, determined by boundary
conditions. The scaling exponents are universal, forming a discrete, strictly
increasing spectrum indexed by the sectors of the symmetry group. When the
initial mean magnetic field is zero, no dynamo effect is found, irrespective of
the anisotropy of the forcing. The rate of isotropization with decreasing
scales
is fully understood from these results.
\end{abstract}

\section{Introduction}
The aims of this paper are two-fold. First, we are interested in the
statistical properties of magnetic fields advected by turbulent velocity
fields. Such magnetic fields possess a ``self-stretching" term that is absent
in the context of advected passive scalars (for a general introduction
see \cite{childress}). Thus a dynamo effect may exist, and its relation to
intermittency and anomalous scaling needs to be addressed. Secondly, we want
to focus on the anisotropic nature of turbulence: generically turbulence is
forced by agents that are neither isotropic nor homogeneous, but most of the
fundamental theories regarding universal scaling properties consider an ideal
model of isotropic turbulence. In the case of a magnetic field advected by a
Gaussian, $\delta$-correlated velocity field with nontrivial spatial scaling
we can present an exact (nonperturbative) solution of the full spectrum of
anomalous scaling exponents of all the anisotropic contributions to the
covariance of the magnetic field. We can thus offer a precise picture of the
rate of isotropization upon diminishing scales, assess the importance of
anisotropy for ``inertial range" scaling, etc.

The equation of motion of a magnetic field $\B.B(\B.r,t)$ reads
\begin{eqnarray}
\partial_t \B.B(\B.r,t)&+&\B.u(\B.r,t)\cdot\B.\nabla \B.B(\B.r,t)
    =\B.B(\B.r,t)\cdot\B.\nabla \B.u(\B.r,t) \nonumber \\
&+&\kappa \nabla^2\B.B(\B.r,t)=\B.f(\B.r,t)
\end{eqnarray}
where $\B.u$ is the advecting velocity field, and $\B.f$ is the external
forcing, and $\kappa$ is the magnetic diffusivity. We address a model in which
the velocity is taken Gaussian, isotropic, $\delta$-correlated in time, and
its correlation function is
\begin{eqnarray}
\langle u^\alpha(\B.r,t)u^\beta(\B.r',t')\rangle
    &=&\delta(t-t')D^{\alpha\beta}(\B.r-\B.r') \nonumber\\
    &=&\delta (t-t')[D^{\alpha\beta}(0)-S^{\alpha\beta}(\B.r)] \ .
\label{eqmot}
\end{eqnarray}
The structure function $\B.S$ scales with exponent $\xi$, $0\le \xi \le 2$:
\begin{equation}
S^{\alpha\beta}(\B.r)=Dr^\xi\left[(\xi+2)\delta^{\alpha\beta}-
\xi\frac{r^\alpha r^\beta} {r^2}\right] \ ,
\quad \lambda \ll r \ll \Lambda \ .
\end{equation}
On the other hand the forcing $\B.f$ is taken here Gaussian,
$\delta$-correlated in time but {\em non-isotropic}. The correlation function
of the forcing has compact support in $\B.k$-space in an interval
$0\le k \le 1/L$, where $L$ is the outer scale of the forcing $\B.f$. We
denote $F^{\alpha\beta}(\B.R)\equiv \langle f^\alpha(\B.R)f^\beta(0)\rangle$.

We are interested in the properties of the covariance of
$\B.B$, $C^{\alpha\beta}(\B.R,t)$,
\begin{equation}
C^{\alpha\beta} (\B.R,t) \equiv
    \langle B^\alpha(\B.R,t)B^\beta(0,t)\rangle \ , \label{C}
\end{equation}
and eventually in the stationary quantity $C^{\alpha\beta} (\B.R)$ which is
obtained in the stationary state if the forcing is balanced by dissipation.
The calculation of this object in an {\em isotropic} ensemble has been
presented by Vergassola \cite{96Ver}. The anisotropic problem has been
addressed recently by Lanotte and Mazzino \cite{99LM}. In the latter study the
covariance (\ref{C}) was not properly expanded in terms of irreducible
representation of the SO(3) symmetry group, and therefore an apparent mixing
of the different sectors was found. As a result the authors had to tackle an
infinite set of equations for all the sectors of the symmetry group. We show
below that this mix-up is spurious, originating from an improper expansion.
Due to the improper choice of the expanding basis set, an infinite linear
coupling between all sectors appeared in \cite{99LM}. In order to solve the
infinite linear system the authors were forced to assume the existence of a
hierarchy between exponents belonging to different sectors and then only a
posteriori to check the correctness of their assumption. In this way the
calculation ends up with one correct set of exponents, as is shown below by
using the proper expansion. We compute additional exponents that were left out
in \cite{99LM}. We will also concern ourselves with the issues of the dynamo
effect and the attainment of a stationary solution for (\ref{C}).

The structure of this paper is as follows: in Sect.2, after presenting the
equations of motion of the covariance, we expand the solutions in terms
of basis functions of the SO(3) symmetry group. In Sect. 3 the above
expansion is used to obtain the matrix representation of the linear
operator which determines the dynamics of the covariance. In Sect. 4
we use this matrix representation to show the absence of a dynamo effect
in the anisotropic sectors of the covariance. Sect. 5 is devoted to the
calculation of the anomalous scaling exponents in the anisotropic sectors,
and Sect. 6 offers a summary and a discussion.
\section{Basic equations and the decomposition in terms of basis functions}

The equation of motion of the covariance were derived by the authors of
\cite{99LM}
with the final result
\begin{eqnarray}
\partial_t C^{\alpha\beta}&=& S^{\mu\nu}\partial_\mu \partial_\nu
    C^{\alpha\beta} -
    [(\partial_\nu S^{\mu\beta})\partial_\mu C^{\alpha\nu} +
    (\partial_\nu S^{\alpha\mu}) \partial_\mu C^{\nu\beta}] \nonumber\\
&+&(\partial_\mu \partial_\nu S^{\alpha\beta}) C^{\mu\nu} +
    2\kappa\nabla^2 C^{\alpha\beta} +
    F^{\alpha\beta} \nonumber\\
&\equiv&{\hat T}^{\alpha\beta}_{~~\sigma\rho}C^{\sigma\rho} +
    F^{\alpha\beta} \ , \label{EqC}\\
\partial_\alpha C^{\alpha\beta} &=& 0 \ , \label{solenoid}
\end{eqnarray}
where the last equation follows from the solenoidal condition for the magnetic
field.

It is advantageous to to decompose the covariance $C^{\alpha\beta}$ in terms
of basis functions that block-diagonalize the angular part of the operator
$\hat \B.T$. These basis functions are implied by the symmetries of
$\hat \B.T$. Since this operator contains only isotropic differential
operators and contractions with either $\delta^{\alpha\beta}$ or
$R^\alpha R^\beta$, it is invariant to all rotations \cite{99ALP}. Accordingly,
the natural basis functions should belong to irreducible representation of the
SO(3) symmetry group, and can be indexed by pairs of indices $j,m$ where
$j=0,1,2,\dots$ and $-j\le m\le j$. We are going to refer to solutions of
(\ref{EqC}) that belong to irreducible representation with a definite $j,m$ as
the ``$j,m$ sector". The operator $\hat \B.T$ leaves such sectors invariant.
In addition, $\hat \B.T$ is invariant to the parity transformation
$\B.R\to -\B.R$, and to the index permutation
$(\alpha,\mu)\Leftrightarrow (\beta,\nu)$. Accordingly, $\hat \B.T$ can be
further block-diagonalized into blocks with definite parity and symmetry under
permutations.

In light of these consideration we seek solutions of the form
\begin{equation}
C^{\alpha\beta}(\B.R,t) =\sum_{q,j,m} a_{q,jm}(R,t) \,\,
    B^{\alpha\beta}_{q,jm}(\hat{\B.R})
\label{c-expand}
\end{equation}
where $\hat{\B.R}\equiv \B.R/R $, and $B^{\alpha\beta}_{q,jm}(\hat{\B.R})$
are tensor functions on the unit sphere, which belong to the sector $j,m$ of
the SO(3) symmetry group. The index $q$ enumerates different tensor functions
belonging to the same sector. While for scalar functions on the sphere there
exist only one spherical harmonic $Y_{jm}$ in each sector, the second rank
tensor functions on the sphere there exist nine different tensors \cite{99ALP}.
The additional symmetries under parity and index permutation group them into
four subgroups with four, two, two and one tensors respectively. With
$\Phi_{jm}( \B.R)\equiv R^j Y_{jm}(\hat {\B.R)}$, in the notation of
\cite{99ALP}, the 4-group (denoted below as subset I) is:
\begin{eqnarray}
B^{\alpha\beta}_{9,jm}(\hat {\B.R})
    &\equiv&R^{-j-2}R^\alpha R^\beta \Phi_{jm}( \B.R)\ , \nonumber\\
B^{\alpha\beta}_{7,jm}(\hat{\B.R})
    &\equiv&R^{-j}(R^\alpha \partial^\beta +R^\beta\partial^\alpha)
    \Phi_{jm}( \B.R) \ , \nonumber\\
B^{\alpha\beta}_{1,jm}(\hat \B.R)
    &\equiv&R^{-j}\delta^{\alpha\beta} \Phi_{jm}( \B.R) \ , \nonumber\\
B^{\alpha\beta}_{5,jm}(\hat \B.R)
    &\equiv&R^{-j+2}\partial^\alpha \partial^\beta \Phi_{jm}( \B.R) \ .
\label{group1}
\end{eqnarray}
These are all symmetric in $\alpha,\beta$ and have a parity of $(-1)^j$.
The 2-groups are denoted respectively as subset II and subset III:
\begin{eqnarray}
B_{8,jm}^{\alpha \beta }(\hat \B.R)
    &\equiv& R^{-j-1}[R^\alpha \epsilon^{\beta\mu\nu} R_\mu\partial_\nu
     + R^\beta \epsilon^{\alpha\mu\nu}R_\mu\partial_\nu] \Phi_{jm}(\B.R)\ , \\
B_{6,jm}^{\alpha\beta}(\hat \B.R)
  &\equiv& R^{-j+1}[\epsilon^{\beta\mu\nu}R_\mu \partial_\nu \partial^\alpha +
                    \epsilon^{\alpha\mu\nu}R_\mu \partial_\nu \partial^\beta]
                    \Phi_{jm}(\B.R) \ .
\end{eqnarray}
\begin{eqnarray}
B_{4,jm}^{\alpha\beta}(\hat \B.R)
    &\equiv& R^{-j-1}\epsilon^{\alpha\beta\mu} R_{\mu }\Phi _{jm}(\B.R) \ , \\
B_{2,jm}^{\alpha\beta}(\hat\B.R)
    &\equiv& R^{-j+1}\epsilon^{\alpha\beta\mu} \partial_\mu \Phi_{jm}(\B.R)
\ .
\end{eqnarray}
The first couple is symmetric to $\alpha,\beta$ exchange and has parity
$(-1)^{j+1}$. The second has the same parity but is antisymmetric to
$\alpha,\beta$ exchange. The remaining basis function is
$B_{3,jm}^{\alpha \beta }(\hat \B.R)
    \equiv R^{-j}(R^\alpha\partial^\beta-R^\beta \partial^\alpha)
    \Phi _{jm}(\B.R)$
which is antisymmetric to $\alpha,\beta$ exchange, with parity $(-1)^j$. This
will be denoted as subset IV. In \cite{99ALP} it was proven that this basis is
complete and indeed transforms under rotations as required for a $j,m$ sector.

It should be noted that not all subsets contribute for every value of $j$. Due
to the obvious symmetry of the covariance:
\begin{equation}
C^{\alpha\beta}(\B.R,t)=C^{\beta\alpha}(-\B.R,t)
\end{equation}
it follows that representations symmetric to $\alpha,~\beta$ exchange must also
have even parity, while antisymmetric representations must have odd parity.
Accordingly, even $j$'s are associated with subsets I and III, and odd $j$'s
are associated with subset II. We show below that subset IV cannot contribute
to this theory due to the solenoidal constraint.

\section{The matrix representation of the operator $\hat {\bf T}$}

Having the angular basis functions we seek the representation of the
operator $\hat \B.T$ in this basis. In such a representation $\hat \B.T$ is a
differential operator with respect to $|\B.R|$ only. In Appendix A we
demonstrate how $\hat \B.T$ mixes basis functions within a given subset, but
not between the subsets - as is expected in the last section. In finding the
matrix representation of $\hat \B.T$ we are aided by the incompressibility
constraint. Consider first subset I with 4 basis functions Eqs.(\ref{group1})
in a given $j,m$ sector. To simplify the notation we will redenote the
coefficients according to $a\equiv a_{9,jm}$, $b\equiv a_{7,jm}$
$c\equiv a_{1,jm}$ $d\equiv a_{5,jm}$. In this basis the operator $\hat \B.T$
takes on the form
\begin{equation}
\hat \B.T \left [\left(
    \begin{array}{c} a\\ b \\ c \\ d \end{array} \right)\right]
=\B. T_1 \left( \begin{array}{c} a''\\ b'' \\ c'' \\ d'' \end{array} \right)
+ \B. T_2\left( \begin{array}{c} a'\\ b' \\ c' \\ d'\end{array} \right)
+ \B. T_3 \left(\begin{array}{c} a\\ b \\ c \\ d \end{array} \right) \ .
\label{T123}
\end{equation}
On the RHS we have matrix products. In addition, the solenoidal condition
implies the following two constrains on $a,~b,~c$ and $d$ (cf. the Appendix of
\cite{99ALP}):
\begin{eqnarray}
0 &=&a' + 2\frac{a}{x} + jb' - j^2\frac{b}{x} + c' - j\frac{c}{x} \nonumber\\
0 &=&b' + 3\frac{b}{x} + \frac{c}{x} + (j-1)d' - (j-1)(j-2)\frac{d}{x}
\end{eqnarray}
Using these conditions one can bring $\B.T_1$ and $\B.T_2$ to diagonal forms,
\begin{eqnarray}
\B.T_1=2(DR^\xi+\kappa) \left(
\begin{array}{cccc}
1 & & & \\
& 1 & & \\
& & 1 & \\
& & & 1
\end{array}
\right) \nonumber\\
\B.T_2=\frac{4}{R}[(DR^\xi+\kappa)+\xi D R^\xi]\left( \begin{array}{cccc} 1
& & & \\
& 1 & & \\
& & 1 & \\
& & & 1
\end{array}
\right) \ .
\end{eqnarray}
$\B.T_3$ can be written in the form
\begin{equation}
\B.T_3 = DR^{\xi-2} \B.Q(j,\xi)+\kappa R^{-2}\B.Q(j,0)
\end{equation}
where the four columns of $\B.Q(j,\xi)$ are
\begin{eqnarray}
\left(
\begin{array}{c}
-(2+\xi )(j+2)(j+3)+2\xi \lbrack (j+1)(2+\xi )+8]+\xi ^{2}(1-\xi )\\ (2+\xi
)(2-\xi )\\ (2+\xi )(2-\xi )(1-\xi )\\0\end{array} \right)\nonumber\\
\left(\begin{array}{c} -2j(j+1-\xi )\xi (2-\xi ) \\ -j(2+\xi )(j+1)+2\xi
(7-\xi )\\ -2j\xi (2+\xi )(2-\xi )\\ 2(2+\xi )(2-\xi
)\end{array}\right)\nonumber\\ \left(\begin{array}{c} -\xi (2-\xi
)(2j-3-\xi )\\ \xi (2-\xi )\\ -j(2+\xi )(j+1)+\xi^{2}(3+\xi )\\
0\end{array}\right)\nonumber\\ \left(\begin{array}{c} -j(j-1)(2-\xi )(4-\xi
)\xi \\ -\xi(j-1)(2-\xi
)(j-4) \\ -j(j-1)(2-\xi )(2+\xi )\xi \\ -(2+\xi )(j-2)(j-1+2\xi )-2\xi
\end{array}
\right)
\end{eqnarray}
In Appendix B we present the two remaining blocks in the matrix representation
of $\hat \B.T$ as a function of $j$.

Next, the single basis $B_{3,jm}$ (subset IV) cannot appear in the theory
since $a_{3,jm}=0$ by the solenoidal condition (cf. Appendix of
\cite{99ALP}):
\begin{eqnarray}
a'_{3,jm} - jR^{-1}a_{3,jm}=0 \ , \\
a'_{3,jm} + R^{-1}a_{3,jm}=0 \ . \nonumber
\end{eqnarray}
Lastly, there are no solutions belonging to the $j=1$ sector. This is due
to the fact
that such solutions correspond to subset II. In this subset the $j=1$
solenoidal condition
implies the equation
\begin{equation}
a'_{8,1m}+\frac{3a_{8,1m}}{R}=0 \ ,
\end{equation}
or $a_{8,1m} \propto R^{-3}$ which is not an admissible solution.

\section{Absence of dynamo effect}
The first issue to clarify is the existence of a stationary solution for
$t\to \infty$. A dynamo effect may cause the covariance to grow unboundedly.
Vergassola \cite{96Ver} showed that this is not the case in the isotropic
sector as long as $\xi<1$. We demonstrate that the dynamo effect is absent
also in the anisotropic sectors. Consider the forceless case of Eq.(\ref{EqC})
with $F^{\alpha\beta}=0$. In addition assume initial conditions such that
$\langle B\rangle=0$. It is easy to see that no mean magnetic field can appear
in time. Accordingly our covariance $C^{\alpha\beta}(\B.R,t)$ tends to zero
when $R\to L$ since $C^{\alpha\beta}(\B.R,t)\to \langle B\rangle^2$. We note
that for $\xi=0$,
$\hat T^{\alpha\beta}_{~~\mu\nu}=
    2\kappa \Delta \delta^\alpha_\mu\delta^\beta_\nu$. In the space of
functions $C^{\alpha\beta}(\B.R,t)$ which vanish outside the domain
$|\B.R|\le L$ this operator is diagonalizable due to its hermiticity, with
negative discrete spectrum $\{-E_\lambda\}$ due to the compactness of the
domain. Thus the general solution in this case is
\begin{equation}
C^{\alpha\beta}(\B.R,t)=\sum_\lambda e^{-E_\lambda t}
C^{\alpha\beta}_\lambda(\B.R) \ .
\end{equation}
In a spherical domain the index $\lambda$ contains the indices $j,m$ and an
index specifying one of the three subsets discussed above. We will assume
that for $\xi\ne 0$ $\hat\B.T$ remains diagonalizable. We will demonstrate
that the eigenvalues $E_\lambda$ remain positive. This will imply that
$C^{\alpha\beta}(\B.R,t)$ is a monotone decreasing function of time, and
hence the absence of a dynamo effect.

To this aim, we define the inner product
\begin{equation}
(\B.C_1,\B.C_2)\equiv\int_{R\le L}
\frac{\left(C_1^{\alpha\beta}\right)^*C_2^{\alpha\beta}}{2(DR^\xi+\kappa)}
d^3R
\end{equation}
and will demonstrate that
\begin{equation}
-E_\lambda (\B.C_\lambda,\B.C_\lambda) =(\B.C_\lambda, \hat \B.T
\B.C_\lambda) <0 \ ,
\label{neg}
\end{equation}
indicating that $E_\lambda >0$. We firstly consider the 4$\times$4 block with
a given $j,m$. In this case $\B.C_\lambda$ is given by
\begin{equation}
\B.C_\lambda (\B.R)=
      a_\lambda(R) \B.B_{9,jm}(\hat\B.R) + b_\lambda(R)\B.B_{7,jm}(\hat\B.R)
    + c_\lambda(R) \B.B_{1,jm}(\hat\B.R)
    + d_\lambda(R)\B.B_{5,jm}(\hat\B.R) \ .
\end{equation}
Using Eq.(\ref{T123}) we obtain
\begin{eqnarray}
(\B.C_\lambda, \hat \B.T \B.C_\lambda)=\int_0^L dR
\frac{R^2}{2(DR^\xi+\kappa)}\left(
a_\lambda^*~~
b^*_\lambda~~c^*_\lambda~~d^*_\lambda\right) \B.M(j) \left[\B. T_1 \left(
\begin{array}{c} a''_\lambda
\\ b''_\lambda \\ c''_\lambda \\ d''_\lambda \end{array} \right) +\B.
T_2\left( \begin{array}{c} a'_\lambda \\ b'_\lambda \\ c'_\lambda \\
d'_\lambda \end{array}
\right)+\B. T_3 \left(
\begin{array}{c} a_\lambda\\ b_\lambda \\ c_\lambda \\ d_\lambda
\end{array} \right)\right] \ ,
\end{eqnarray}
where the matrix $\B.M(j)$ arises from the angular integration over the
spherical tensors $\B.B_{q,jm}$. This matrix is obtained by a direct
calculation. For example
$M_{1,1}(j)\equiv\int d\hat\B.R \B.B^*_{9,jm}(\hat\B.R)\B.B_{9,jm}(\hat\B.R)$.
The full matrix reads
\begin{eqnarray}
\B.M(j)=
\left(\begin{array}{cccc}
    1 & 2j & 1 & j(j-1) \\
    2j & 2j(3j+1) & 2j & 2j(j-1)(2j+1) \\
    1 & 2j & 3 & 0 \\
    j(j-1) & 2j(j-1)(2j+1) & 0 & j(j-1)(2j-1)(2j+1)
\end{array} \right)
\end{eqnarray}
We note that $\B.M(j)$ is symmetric and positive definite. By integration
by parts, using the fact that our covariances vanish for $R=L$, we demonstrate
in Appendix C that Eq.(\ref{neg}) is true.

One important conclusion of this calculation is the relative rate of decay of
the various anisotropic contributions. We see that upon increasing $j$ the
inner product (\ref{neg}) becomes more negative. Thus, any anisotropic initial
conditions results in a rapid decay of the higher $j$ contributions. Without
anisotropic forcing the covariance of the magnetic field becomes isotropic in
time. We will show below that also in the (anisotropic) stationary state
maintained by anisotropic forcing, the covariance isotropizes on the smaller
scales. The scaling exponents governing the $R$ dependence are also strictly
increasing with increasing $j$. Thus invariably for small enough scales and
for long times one restores local isotropy.

\section{calculation of the scaling exponents}

In the absence of a dynamo effect, we can consider a stationary state of the
system, maintained by the forcing term $\B.f(\B.r,t)$. The covariance in such
a case will bey the following equation:
\begin{equation}
0={\hat T}^{\alpha\beta}_{~~\sigma\rho}C^{\sigma\rho}+F^{\alpha\beta} \ .
\label{EqT}
\end{equation}
Deep in the inertial range we look for scale invariant solutions, obtained as
zero-modes of Eq.(\ref{EqT}). Indeed, when $\xi>0$ and well within the
inertial range we can take the magnetic dissipation to zero, and as a result,
the homogeneous part of Eq.(\ref{EqT}) (without $F^{\alpha\beta}$) will be
scale invariant, leading to scale invariant solutions. We will need to match
these zero modes to the appropriate zero modes computed in the dissipative
range at the end. This will necessitate the discussion of zero modes when
$\xi=0$, and see below.

The calculation of the scale-invariant solutions becomes rather immediate once
we know the functional form of the operator $\hat{\B.T}$ in the basis of the
angular tensors $\B.B_{q,jm}$. Using the expansion (\ref{c-expand}), and the
fact that $\hat{\B.T}$ is block diagonalized by such an expansion, we get a
set of 2nd order coupled ODE's for each block. To demonstrate this point,
consider the four dimensional block of $\hat{\B.T}$, created by the four basis
tensors $\B.B_{q,jm}$ of subset I. According to the notation of the last
section, we denote the coefficients of these angular tensors in
(\ref{c-expand}), by the four functions $a(R),b(R),c(R),d(R)$:
\begin{equation}
C^{\alpha\beta}(\B.R) \equiv a(R)B^{\alpha\beta}_{9,jm} +
b(R)B^{\alpha\beta}_{7,jm} + c(R)B^{\alpha\beta}_{1,jm} +
d(R)B^{\alpha\beta}_{5,jm} + \dots \ ,
\end{equation}
where $\dots$ stand for terms with other $j,m$ and other symmetries with
the same $j,m$. According to (\ref{T123}), well within the inertial range,
these functions obey:
\begin{equation}
\B. T_1(\kappa=0)
\left(\begin{array}{c}a''\\ b'' \\ c'' \\ d'' \end{array} \right) +
\B.T_2(\kappa=0)\left(\begin{array}{c} a'\\ b'\\ c'\\ d'\end{array}\right) +
\B.T_3(\kappa=0) \left( \begin{array}{c} a\\ b\\ c\\ d \end{array}\right)=0 \ .
\label{zero-eq}
\end{equation}
Due to the scale-invariance of these equations, we look for scale-invariant
solutions in the form:
\begin{equation}
a(R)=aR^{\zeta}, \quad b(R)=bR^{\zeta},\quad d(R)=cR^{\zeta},
\quad d(R)=dR^{\zeta} \ .
\label{si-form}
\end{equation}
Substituting (\ref{si-form}) into (\ref{zero-eq}) results in a set of four
linear homogeneous equations for the unknowns $a,b,c,d$ :
\begin{equation}
\left[
  \zeta(\zeta-1) \B.T_1(\kappa=0) + \zeta \B.T_2(\kappa=0) + \B.T_3(\kappa=0)
\right]
\left(\begin{array}{c} a\\ b \\ c \\ d \end{array} \right) =0 \ .
\label{zero-linear-eq}
\end{equation}
The last equation admits non-trivial solutions only when
\begin{equation}
\det\left[
  \zeta(\zeta-1) \B.T_1(\kappa=0) + \zeta \B.T_2(\kappa=0) + \B.T_3(\kappa=0)
\right]=0 \ .
\end{equation}
This solvability condition allows us to express $\zeta$ as a function of $j$
and $\xi$. Using MATHEMATICA we find eight possible values of $\zeta$,
out-of-which, only four are in agreement with the solenoidal condition:
\begin{eqnarray}
\zeta^{(j)}_{i}&=&-\frac{1}{2}\xi -\frac{3}{2} \pm \frac{1}{2}
    \sqrt{H(\xi ,j) \pm 2 \sqrt{K(\xi ,j)}} \label{exponents-I} \\
K(\xi,j) &\equiv& \xi^4- 2\xi^3 + 2\xi^3j + 2\xi^3j^2 - 4\xi^2j -
    3\xi^2 - 4\xi^2 j^2 - 8 \xi j^2-8\xi j + 4\xi + 16j + 16j^2 + 4
    \nonumber \\
H(\xi ,j) &\equiv& -\xi^2 - 8\xi + 2\xi j^2 + 2\xi j + 4j^2 + 4j + 5
\nonumber \ .
\end{eqnarray}
Not all of these solutions are physically acceptable because not all of them
can be matched to the zero mode solutions in the dissipative regime. To see
why this is so, consider the zero-mode equation deep inside the dissipative
regime. Here the dissipation terms become dominant and we can neglect all
other terms in $\hat{\B.T}$. The zero mode equation in this regime becomes
a simple Laplace equation:
\begin{equation}
    2\kappa \nabla^2 \B.C = 0 \ .
\end{equation}
Notice however, that up to an overall factor of $\kappa / (\kappa+D)$, this
equation is identical to the zero-mode equation for the special case $\xi=0$.
Accordingly, the solutions $a(R),b(R),c(R),d(R)$ are scale invariant with the
exponents $\zeta^{(j)}_{i}|_{\xi=0}$. Using this we can immediately
distinguish between the physical and unphysical solutions: At $R=0$ the
covariance must be finite, and therefore only the positive exponents in
(\ref{exponents-I}) are acceptable. Let us now consider the zero mode
equation for $\xi=0$. In this particular case, the equation is
\begin{equation}
    (2\kappa+2D) \nabla^2 \B.C = 0
\end{equation}
throughout the whole range, and so the zero modes do not change their
functional form as we move from the dissipative regime to the inertial regime.
Therefore for, $\xi=0$, also in the inertial regime, only the positive
exponents are legitimate. Assuming now that the solution (including the
exponents) are continuous in $\xi$, (and not necessarily analytic!), we
immediately get, that also for finite $\xi$, only the positive exponents
appear in the inertial range. Finally there exist two branches of solutions
corresponding to the (-) and (+) in the square root.
\begin{equation}
\zeta^{(j)}_{I\pm}=-\frac{3}{2}-\frac{1}{2}\xi + \frac{1}{2} \sqrt{H(\xi ,j)
    \pm 2 \sqrt{K(\xi ,j)}}\ , \quad \text {subset I}
\end{equation}
These exponents are in agreement with \cite{99LM}. Note that in the case of
$j=0$
only $\zeta^{(0)}_{I+}$ exists since the other exponent is not admissible,
being
negative for $\xi\to 0$, and therefore excluded by continuity. For
$j\ge 2$ both solutions are admissible, and the leading one is the negative
exponent
which is smaller.

In addition to these one
needs to compute the exponents corresponding to the subsets II and III. The
computation in the other two blocks follows the same lines. Since these are
2$\times$2 they furnish two solutions for the exponents, one of which is
negative. We end up finding
\begin{equation}
\zeta^{(j)} _{II}=-\frac{3}{2}-\frac{1}{2}\xi +\frac{1}{2}
    \sqrt{1-10\xi+ \xi ^{2}+2j^{2}\xi +2j\xi +4j+4j^{2}}\ ,
    \quad \text {subset II}
\end{equation}
\begin{equation}
\zeta^{(j)} _{III}=-\frac{3}{2}-\frac{1}{2}\xi +\frac{1}{2}\sqrt{ \xi ^{2}
    + 2\xi + 1 + 4j^{2} + 2j^{2}\xi + 4j + 2\xi j}\ ,
    \quad \text {subset III}
\end{equation}
For $j=0$ there is no contribution from this subset, as the exponent is
negative. The dependence of the admissible leading exponents on $\xi$
is displayed in Fig.1 and Fig.2.
In Table 1 we summarize which are the leading exponents in each sector.

In addition to matching the zero modes to the dissipative range, one has to
guarantee matching at the outer scale $L$. The condition to be fulfilled is
that the sum of the zero-modes with the inhomogeneous solutions (whose
exponents are 2-$\xi$) must give $\B.C(\B.R)\to 0$ as $|\B.R|\to L$. Obviously
this means that the forcing must have a projection on any sector $\B.B_{q,jm}$
for which $a_{q,jm}$ is nonzero.

\section{Summary and conclusions}

The results of this paper should be examined in light of the recent progress
in understanding the effects of anisotropy on the statistics of fully
developed turbulence \cite{99ALP,98ADKLPS,99ABMP,99KLPS}. Whereas in the
Navier-Stokes case one cannot present exact results, the present study can
affords exact calculations of the whole spectrum of scaling exponents that
determines the covariance of a vector field in the presence of anisotropy. We
have presented a detailed and systematic investigation of scaling properties
of the covariance of a magnetic field advected by a gaussian and
delta-correlated in time velocity field. We have extended the non-perturbative
analysis presented by Vergassola in \cite{96Ver} for the isotropic sector to
all the sectors of the SO(3) symmetry group. Our analysis leads to the
conclusions that the scaling exponents are strictly increasing with the index
of $j$ of the sector, meaning that there is a tendency toward isotropization
upon decreasing the scales of observation. We also showed that as far as the
dynamo problem is concerned, anisotropic sectors are less unstable than the
isotropic sector: in the absence of an external forcing anisotropies decay in
time faster then isotropic fluctuations. In distinction with the expansion
presented in \cite{99LM}, our results are free of any assumptions about the
hierarchy of scaling exponents belonging to different SO(3) sectors. This is
due to the employment of a proper basis set. The equations for the magnetic
covariance foliate into independent closed equations for each set of
irreducible
representations of the SO(3) group.

In summary, we have shown that the covariance of the magnetic field is
naturally computed as a sum of contributions proportional to the irreducible
representations of the SO(3) symmetry group. The amplitudes are non-universal,
determined by boundary conditions. The scaling exponents are universal,
forming a discrete, strictly increasing spectrum indexed by the sectors of the
symmetry group. Similar results were presented for passive scalar fluctuations
in \cite{99ALPP}, and for Navier-stokes fluctuations in
\cite{99ALP,98ADKLPS,99ABMP,99KLPS}. In the present case anomalous scaling
laws are found as the zero-modes of the inertial operator governing the
stationary equation for the magnetic covariance \cite{GK,95CFKL}. Matching
with the UV boundary conditions selects the physically acceptable solutions.
It appears quite clear now that the issue of anomalous, universal scaling
exponents in turbulence has ramified to the multitude of sectors of the
appropriate symmetry groups.

\acknowledgments
We are grateful to M. Vergassola for many useful discussions and suggestions.
We also thank A. Lanotte and A. Mazzino for a fruitful exchange of ideas. At
Weizmann this work has been supported in part by the German Israeli Foundation,
the European Commission under the Training and Mobility of Researchers program,
and the Naftali and Anna Backenroth-Bronicki Fund for Research in Chaos and
Complexity. In Rome this research has been partially supported by
INFM (PRA-TURBO) and by the EU contract FMRX-CT98-0175.

\appendix
\section{Demonstration of the action of ${\hat T}^{\alpha\beta}_{~~\mu\nu}$}

As an example of the operation of $\hat \B.T$ on the basis function, consider
an explicit calculation of $\partial^2 C^{\alpha\beta}$. Such a term appears
as a part of $S^{\mu\nu}\partial^\mu \partial^\nu$ which is a part of
$\hat\B.T$, and also in the magnetic dissipation term. Considering explicitly
the the part $a_{9,jm}(R,t)B_{9,jm}^{\alpha\beta}(\hat\B.R)$:
\begin{eqnarray}
&&\partial^2 a_{9,jm} R^{-j-2}R^\alpha R^\beta \Phi_{jm} =
    \partial^\mu \partial_\mu a_{9,jm} R^{-j-2} R^\alpha R^\beta
    \Phi_{jm}\nonumber \\
&& = \partial^\mu [a'_{9,jm} R^{-j-3}-(j+2)a_{9,jm}R^{-j-4}] R_\mu R^ \alpha
    R^\beta \Phi_{jm} \nonumber\\
&& + \partial^\alpha a_{9,jm}R^{-j-2}R^\beta \Phi_{jm} +\partial^\beta
    a_{9,jm}R^{-j-2}R^\alpha \Phi_{jm} \nonumber\\
&& + \partial^\mu a_{9,jm} R^{-j-2}R^\alpha R^\beta \partial_\mu
    \Phi_{jm}\nonumber \\
& = & [a''_{9,jm} -(j+3)\frac{a'_{9,jm}}{R} - (j+2)\frac{a'_{9,jm}}{R}
    +(j+2)(j+4)\frac{a_{9,jm}}{R^2} ]
    B^{\alpha\beta}_{9,jm} \nonumber\\
&& + (j+5)[ \frac{a'_{9,jm}}{R} - (j+2) \frac{a_{9,jm}}{R^2} ]
    B^{\alpha\beta}_{9,jm} \nonumber \\
&& + 2[ \frac{a'_{9,jm}}{R} -(j+2)\frac{a_{9,jm}}{R^2}]
    B^{\alpha\beta}_{9,jm} + 2\frac{a_{9,jm}}{R^2}
    B^{\alpha\beta}_{1,jm} + \frac{a_{9,jm}}{R^2}B^{\alpha\beta}_{7,jm}
    \nonumber\\
&& + j[\frac{a'_{9,jm}}{R} - (j+2)\frac{a_{9,jm}}{R^2}]B^{\alpha\beta}_{9,jm}
    + \frac{a_{9,jm}}{R^{2}} B^{\alpha\beta}_{7,jm} \nonumber\\
&& = [a''_{9,jm} + 2\frac{a'_{9,jm}}{R} -(j+2)(j+3)
    \frac{a_{9,jm}}{R^2}] B^{\alpha\beta}_{9,jm} +2\frac{a_{9,jm}}{R^2}
    B^{\alpha\beta}_{7,jm} +2\frac{a}{R^{2}}B^{\alpha\beta}_{1,jm}
\end{eqnarray}
In performing the computation we make use of the following basic identities
that are employed repeatedly in all our calculations:
\begin{eqnarray}
\partial^\mu \partial_\mu \Phi_{jm} =0 \ , \\
R^\mu \partial_\mu \Phi_{jm} = j \Phi_{jm}
\end{eqnarray}
The first identity follows from $\partial^2 Y_{jm}= -j(j+1)R^{-2} Y_{jm}$.
The second from the fact that $\Phi_{jm}$ are homogeneous polynomials of
degree $j$. As expected, the result remains in a $j,m$ sector, and mixes only
basis functions with the same symmetry properties.

\section{$\hat {\bf T}$ and the solenoidal condition in the two remaining
         subsets}

In this Appendix we present the two blocks pertaining to the $(-1)^{j+1}$
parity. The part denoted in Eq.(\ref{T123}) as $\B.T_1$ and $\B.T_2$ remains
unchanged except that the identity matrix is now 2-dimensional. For the case
of invariance under $\alpha,\beta$ interchange (subset II) we find the
2$\times$2 matrix $\B.Q(j,\xi)$:
\begin{eqnarray}
\left(\begin{array}{cc}
-(j+1)(2+\xi )(j+2-\xi )+2\xi (7-\xi ) & -\xi (j-1)^{2}(2-\xi ) \\
(2-\xi)(2+\xi ) & j(j-1)(2+\xi )+\xi (j-3)(2+\xi )+2\xi
\end{array}\right) \ .
\end{eqnarray}
The solenoidal condition reads in this case (cf. Appendix of \cite{99ALP}):
\begin{equation}
a'_{8,jm}+3R^{-1}a_{8,jm}+(j-1)a'_{6,jm}-(j-1)^2R^{-1}a_{6,jm}=0 \ .
\end{equation}
>From this equation we learn that a contribution pertaining to $j=1$ cannot
appear in this theory, since for this value of $j$ $a'_{8,jm}$ must have a
negative scaling exponent which is not admissible.

For the case of antisymmetry under $\alpha,\beta$ interchange (subset III) we
find the 2$\times$2 matrix $\B.Q(j,\xi)$:
\begin{eqnarray}
\left(
\begin{array}{cc}
\xi (4+2\xi +4j)-(j+1)(j+2)(2+\xi ) & \xi j(j-1)(2-\xi ) \\
4-2\xi & -(j-1)[j(2+\xi )+4\xi ]
\end{array}
\right) \ ,
\end{eqnarray}
with the solenoidal condition (cf. Appendix of \cite{99ALP}:
\begin{equation}
R^{-1} a_{4,jm} -a'_{2,jm} +(j-1)R^{-1}a_{2,jm}=0 \ .
\end{equation}

\section{Proof of Eq.(19)}

To demonstrate Eq.(\ref{neg}) we note $\hat \B.T$ as well as $\B.M(j)$ are
$m$ independent. We can therefore consider without loss of generality the
$m=0$ case. In this case the basis functions as well as the coefficients
$a$, $b$, $c$ and $d$ are real. For non-zero $m$ the imaginary components have
to cancel with the imaginary components of $-m$ since the covariance is real.
We treat separately the contributions associated with $\B.T_{1},\B.T_{2}$ and
with $\B.T_3$. showing that they are all negative definite.

\subsection{The integrals of ${\bf T}_1$ and ${\bf T}_2$.}
For the evaluation of these integrals it is convenient to work in the basis
that diagonalizes $\B.M(j)$. Since $\B.M(j)$ is a real and symmetric matrix,
it is diagonalizable and as it is non-negative its eigenvalues $\mu_i$,
$i=1,~2,~3,~4$, are nonnegative. $T_{1},T_{2}$ are proportional to the unit
matrix, and therefore they remain so in any basis, and in particular in the
diagonal basis of $M$. In that basis, $a,b,c,d$ are replaced by
$a_1,a_2,a_3,a_4$, and the contribution of $T_{1},T_{2}$ is:
\begin{equation}
\sum_{i=1}^4 \mu_i \int\limits_{0}^{L} dx \frac{x^{2}}{Dx^\xi+\kappa}
    a_i [2(Dx^\xi + \kappa) a''_i + 4(Dx^\xi +\kappa)\frac{a'_i}{x}
    + 4\xi Dx^\xi \frac{a'_i}{x}]
\end{equation}
This integral is negative definite for all values of $i$ since it is the sum
of two negative definite integrals $I_1$ and $I_2$:
\begin{eqnarray}
I_1 &=& \int\limits_{0}^{L}dx \frac{x^2}{Dx^\xi +\kappa}
    a_i[2(Dx^\xi + \kappa) a''_i + 4(Dx^\xi + \kappa)
    \frac{a'_i}{x}] \nonumber\\
&=& 2\int\limits_{0}^{L}dx (x a_i)\frac{d^2}{dx^2}(x a_i) \nonumber\\
&=& -2\int\limits_{0}^{L}dx \left[\frac{d}{dx}(x a_i)\right]^2 < 0 \\
I_2 &=& 4D\int\limits_{0}^{L}dx \frac{x^2}{Dx^\xi +\kappa} a_i x^\xi
    \frac{a'_i}{x} \nonumber\\
&=& -4D\int\limits_{0}^{L}dx \frac{d}{dx} \left[ \frac{x^{\xi+1}}
    {Dx^\xi+\kappa}\right] a_i^2 - I_2
\end{eqnarray}
Accordingly,
\begin{eqnarray}
I_2 &=& -2D\int\limits_{0}^{L}dx \frac{d}{dx} \left[\frac{x^{\xi+1}}
    {Dx^\xi+\kappa}\right] a_i^2 \nonumber\\
&=& -2D\int\limits_{0}^{L}dx \frac{Dx^{2\xi} +\kappa(1+\xi) x^\xi}
    {\left(Dx^\xi + \kappa\right)^2 }
    a_i^2 < 0 \ .
\end{eqnarray}

\subsection{The integral of ${\bf T}_3$}
The contribution of $\B.T_{3}$ has two parts: One which is proportional to
$\kappa$, and one which is proportional to $D$. We shall analyze each of
them separately and show that $\B.M(j)\cdot \B.T_{3}$ is a nonpositive matrix
for every $j\geq 2$ and every $0\leq \xi \leq 2$.

\begin{enumerate}
\item The part involving $\kappa $ is:
\begin{eqnarray}
I_3 &=& \int\limits_{0}^{L}dx \frac{x^2}{Dx^\xi +\kappa}
    \left(\begin{array}{cccc} a & b & c & d \end{array}\right)
    \B.M(j) \kappa x^{-2}\B.Q(j,0)
    \left( \begin{array}{c} a \\ b \\ c \\ d \end{array} \right) \\
&=& \kappa \int\limits_{0}^{L}dx \frac{1}{Dx^\xi +\kappa}
    \left(\begin{array}{cccc} a & b & c & d \end{array} \right)
    \frac{1}{2}[\B.M(j) \B.Q(j,0) +\left(\B.M(j) \B.Q(j,0) \right)^{T}]
    \left( \begin{array}{c} a \\ b \\ c \\ d \end{array} \right) \\
&\equiv& \kappa \int\limits_{0}^{L}dx \frac{1}{Dx^\xi + \kappa}
    \left(\begin{array}{cccc}a & b & c & d\end{array}\right) \B.X(j,0)
    \left(\begin{array}{c}a \\b \\c \\d \end{array}\right) \ .
\end{eqnarray}
where $\B.X(j,\xi)$ is the symmetric matrix
\begin{equation}
\B.X(j,\xi) \equiv \frac{\B.M(j)\B.Q(j,\xi)+(\B.M(j)\B.Q(j,\xi))^T}{2} \ .
\end{equation}
For $j=2, \xi=0$, $\B.X(j,\xi)$ is given by
\begin{equation}
\B.X(2,0) = \left(\begin{array}{cccc}
-20 & -32 & -12 & 0 \\
-32 & -176 & -48 & 0 \\
-12 & -48 & -36 & 0 \\
0 & 0 & 0 & 0
\end{array}\right) \ ,
\end{equation}
with the eigenvalues $(-12.97..,-196.43..,-21.59..,0)$ so the expression is
obviously non-positive. For higher $j$'s, we can look at the determinant of
$X(j,0)$:
\begin{equation}
    \det \B.X(j,0)=(j+3)(j+2)^2 (j+1)^4 j^4 (j-1)^2 (j-2)
\end{equation}
this function is positive for every $j>2$ which means that we have 4 negative
eigenvalues when $j>2$.

\item The part involving $D$ is:
\begin{eqnarray}
I &=& D\int\limits_{0}^{L}dx \frac{x^2}{Dx^\xi +\kappa}
    \left(\begin{array}{cccc} a & b & c & d \end{array}\right)
    \B.M(j)x^{\xi-2}\B.Q(j,\xi)
    \left(\begin{array}{c} a \\ b \\ c \\ d \end{array} \right) \nonumber\\
&=&D\int\limits_{0}^{L}dx \frac{x^\xi}{Dx^\xi + \kappa}
    \left(\begin{array}{cccc}a & b & c & d\end{array}\right)
    \B.X(j,\xi )\left(\begin{array}{c} a \\ b \\ c \\ d \end{array}\right)
\end{eqnarray}
The proof of the nonpositivity of this expression follows the same lines of
the previous discussion. We know that for $\xi=0,j=2$ $\B.X(j,\xi)$ has three
negative eigenvalues and one zero. Therefore, It is sufficient to show that
$\det \B.X(j,\xi)$ is positive for every $0<\xi<2$ and $j \ge 2$ to ensure
that $\B.X(j,\xi)$ is indeed nonpositive. This is indeed the case, as can be
verified explicitly using MATHEMATICA.
\end{enumerate}

\begin{table}
\begin{tabular}{|l|l|l|}
~& Symmetric~~~~~~~~~~ & AntiSymmetric~~~~~~~~~~~ \\ \hline\hline
$j=0$ &$\zeta _{I+}$ & - \\
Even $j>0$ & $\zeta _{I-}$ &$\zeta _{III}$ \\
Odd $j>1$ & $\zeta _{II}$ & - \\
\end{tabular}
\caption{The leading exponents in the various sectors}
\end{table}
\begin{figure}
\epsfxsize=8.0truecm
\epsfysize=6.0truecm
\epsfbox{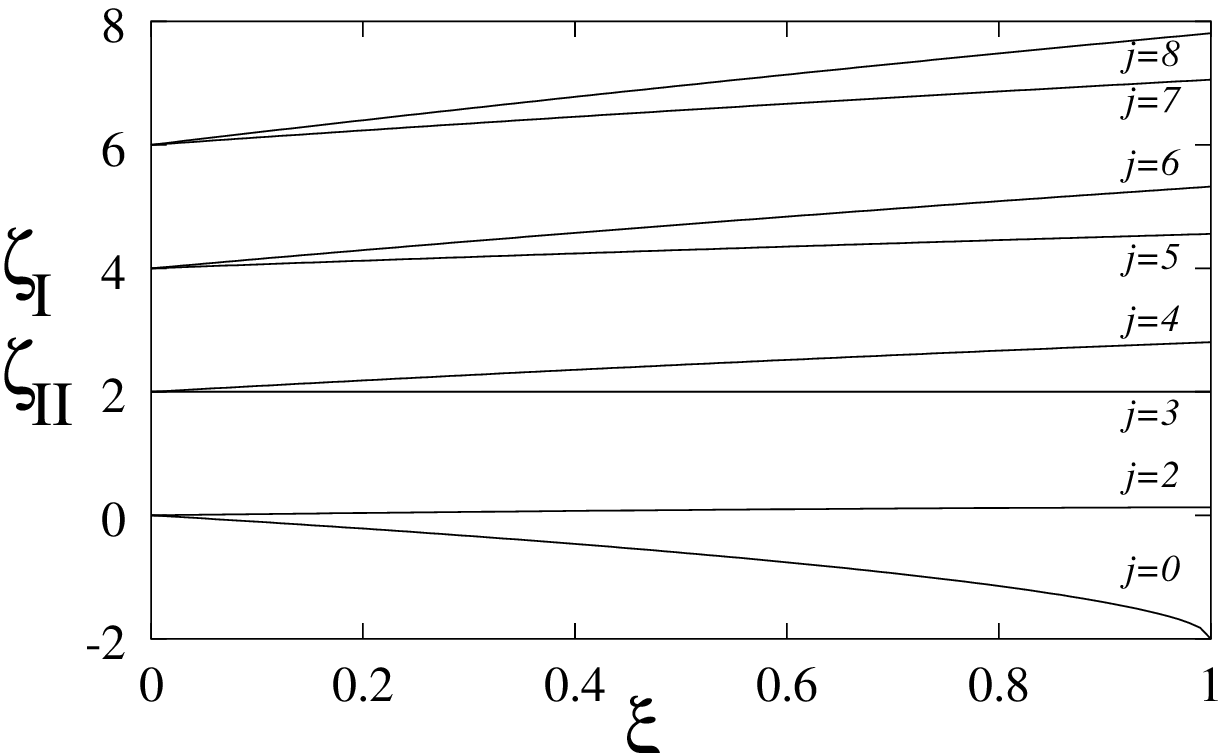}
\caption{The leading exponents of the symmetric parts
of the zero modes of the magnetic covariance}
\label{exponent}
\end{figure}
\begin{figure}
\epsfxsize=8.0truecm
\epsfysize=6.0truecm
\epsfbox{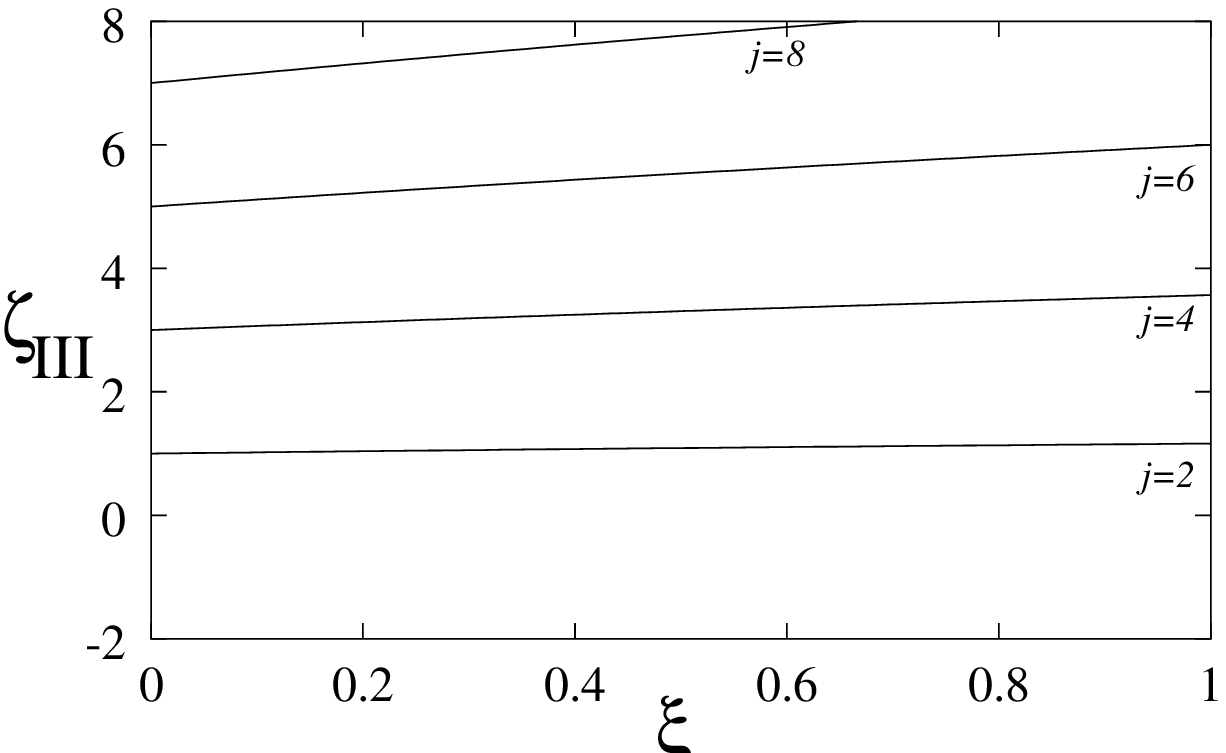}
\caption{The leading exponents of the antisymmetric parts
of the zero modes of the magnetic covariance}
\label{exponents}
\end{figure}

\begin{references}

\bibitem{childress}
S. Childress and A. Gilbert, {\em Twist Stretch Fold: the Fast Dynamo}
(Springer, Berlin, 1995).

\bibitem{96Ver}
M. Vergassola, Phys. Rev E, {\bf 53}, R3021 (1996).

\bibitem{99LM}
A. Lanotte and A. Mazzino, e-print archive LANL, {\em chao-dyn/9903026}

\bibitem{99ALP}
I.Arad, V.S. L'vov and I. Procaccia, Phys. Rev. E {\bf 59}, 6753 (1999).

\bibitem{99ALPP}
I. Arad, V.S. L'vov, E. Podivilov and I. Procaccia , "Anomalous Scaling in the
Anisotropic Sectors of the Kraichnan Model of Passive Scalar Advection",
Phys Rev. E, submitted.

\bibitem{GK}
K. Gawedzki and A. Kupiainen, Phys. Rev. Lett., {\bf 75}, 3834 (1995).

\bibitem{95CFKL}
M. Chertkov, G. Falkovich, I. Kolokolov and V. Lebedeev, Phys. Rev E {\bf 52}
4924 (1995)

\bibitem{98ADKLPS}
I. Arad, B. Dhruva, S. Kurien,  V.S. L'vov, I. Procaccia and K.R.Sreenivasan,
Phys. Rev. Lett., {\bf 81}, 5330
(1998)

\bibitem{99ABMP}
I. Arad, L. Biferale, I. Mazzitelli and I. Procaccia, Phys. Rev. Lett.
{\bf 82}, 5040 (1999).

\bibitem{99KLPS}
S. Kurien, V. S. L'vov, I. Procaccia and  K.R. Sreenivasan, Phys. Rev. E,
in press.

\end{references}
\end{document}